\documentclass[final,3p,times]{elsarticle}

\usepackage{amssymb}
\usepackage{amsmath} 

\usepackage{siunitx}

\usepackage[version=3]{mhchem} 
\usepackage{chemarr}

\journal{Physica A}
       
\begin{document}

\begin{frontmatter}
 
\title{On the Gouy-Chapman-Stern model of the electrical double-layer structure with a generalized Boltzmann factor}

\author[sree,camr,fiu]{Anis Allagui\corref{cor1}}
\cortext[cor1]{Corresponding author} 
\ead{aallagui@sharjah.ac.ae}
\address[sree]{Dept. of Sustainable and Renewable Energy Engineering, University of Sharjah, Sharjah, UAE}
\address[camr]{Center for Advanced Materials Research, Research Institute of Sciences and Engineering, University of Sharjah, Sharjah, UAE}
\address[fiu]{Dept. of Mechanical and Materials Engineering, Florida International University, Miami, FL33174, United States}

\author[phys]{Hachemi Benaoum} 
\address[phys]{
Dept. of Applied Physics and Astronomy, 
University of Sharjah, PO Box 27272, Sharjah, United Arab Emirates 
}

\author{Oleg Olendski} 
\address[phys]{
Dept. of Applied Physics and Astronomy, 
University of Sharjah, PO Box 27272, Sharjah, United Arab Emirates 
}

\begin{abstract}

The classical treatment of the electrical double-layer (EDL) structure at a planar metal/electrolyte junction via the Gouy-Chapman-Stern (GCS) approach  is based on the Poisson equation relating the  electrostatic potential to the net mean charge density.  The  ions  concentration in the diffuse layer are assumed to follow the Boltzmann distribution law, i.e. $\propto \exp(-\tilde{\psi})$ where $\tilde{\psi}$ is the dimensionless electrostatic potential. However, even in stationary equilibrium in which variables are averaged over a large number of elementary stochastic events, deviations from the mean-value are expected. In this study we evaluate the behavior of the EDL by assuming some small perturbations superposed on top of its Boltzmann  distribution of ion concentrations using the Tsallis nonextensive statistics framework. With this we assume the ion concentrations to be proportional to $  [ 1- (1-q) \tilde{\psi}]^{{1}/{(1-q)}} = \exp_q({- \tilde{\psi}})$ with $q$ being a real parameter that characterizes the system's statistics. We derive analytical expression and provide computational results for the overall differential capacitance of the EDL structure, which,  depending on the values of the parameter $q$ can 
show both the traditional inverse bell-shaped curves for aqueous solutions and  bell curves observed with ionic liquids. 

\end{abstract}


  \end{frontmatter}

\section{Introduction}

\begin{figure}[h]
\begin{center}
 \includegraphics[width=3.2in]{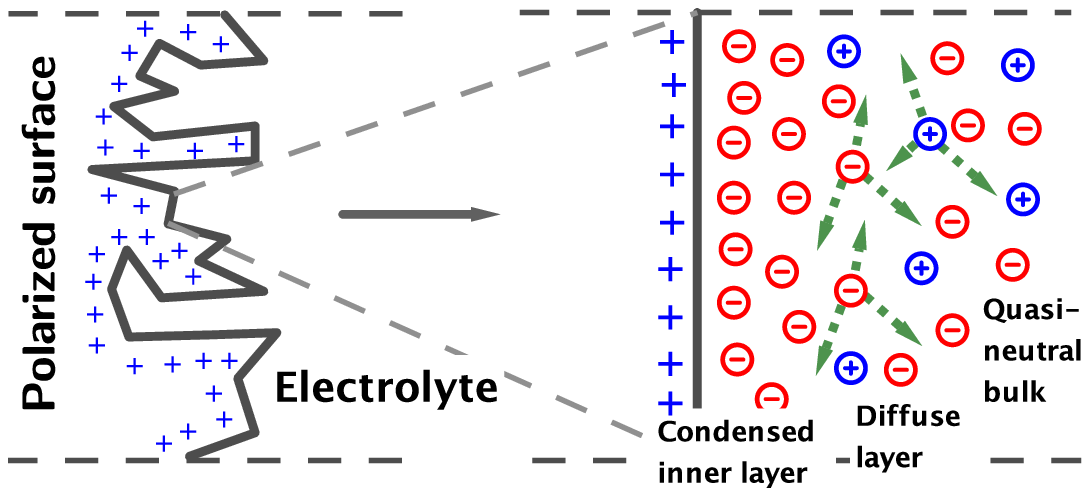}
\caption{Schematic representation of electric double-layer structure at a positively-polarized blocking electrode/electrolyte interface. Its thickness (a measure of its capacity) is a  usually in the order of a few nanometers at most.}
 \label{fig1}
\end{center}
\end{figure}

The electrical double-layer (EDL) is more than a century-old fundamental topic of electrochemistry and physical chemistry \cite{delahay1965double, doi:10.1002/9780470142806.ch2, parsons1990electrical, bazant2004diffuse, BiesheuvelandDykstra}. 
It is found in many applications such as in membrane science and technology \cite{mclaughlin1989electrostatic, moy2000tests}, 
water desalination by capacitive ion removal \cite{biesheuvel2010nonlinear, hartel2015fundamental, BiesheuvelandDykstra}, capacitive mixing \cite{brogioli2009extracting, rica2012thermodynamic, simoncelli2018blue}, energy storage in porous electrodes and fractional-order capacitors \cite{biesheuvel2011diffuse, bazant2004diffuse, foedlc, EC2015, memoryAPL, memQ}, colloidal suspensions  \cite{hansen2000effective, brown2016determination}, etc.
It refers to the general phenomenon of charge separation across an interfacial surface, which can be imagined as a plane that separates two adjacent and distinct phases (e.g. metal and electrolyte) having excess electronic or ionic charges (see Fig.\;\ref{fig1}).    
At such an electrified interface, the electrode charge is not necessarily  constant, but on the contrary fluctuates in response to the stochastic thermal motions in the adjacent electrolyte. These fluctuations are related to a set of configuration variables including the number, positions and momenta of mobile ions of each species near the electrode, and become important as the correlation between species increases \cite{outhwaite1974higher, uralcan2016concentration}. 
The extent of these correlations is related to the nature of the interfacial liquid, and is expected to increase with the increase of ions concentration as well as in nanometer-sized confined geometries at the electrode surface. 
As a result the electrical properties of within small scales of the electrode, such as the differential capacitance, electrostatic potential or accumulated charge density for instance, 
 are expected to be affected by the statistics  of these microscopic fluctuations\;\cite{limmer2013charge, uralcan2016concentration}.

Several computational models such as 
 Monte-Carlo (MC), molecular dynamics (MD) 
 and density functional theory (DFT) simulations are now very apt to provide good descriptions of the EDL structure\;\cite{hartel2015fundamental, limmer2013charge, simoncelli2018blue, hartel2017structure}. However, because of the computational resources needed for the explicit microscopic analysis of large numbers of interacting solvent and mobile ions, mean-field calculations remain popular due their simplicity and comparatively good accuracy\;\cite{BiesheuvelandDykstra, biesheuvel2011diffuse, bazant2004diffuse}.  
 The classical Gouy-Chapman-Stern (GCS) approach in which  the Poisson-Boltzmann (PB)  equations are used for the treatment of the diffuse part of the solvent (reviewed in Sect.\;1) is widely used because it is verified to be asymptotically correct in the weak-coupling regime and  it provides the essential of the EDL   properties\;\cite{OLDHAM2008131, delahay1965double, gray2018nonlinear, BiesheuvelandDykstra, moy2000tests, corry2000tests, doi:10.1002/9780470142806.ch2}. It  assumes the equilibrium ion concentrations to decay exponentially with the distance from the interface following a  Boltzmann distribution function with respect to the mean electrostatic potential energy in the EDL. This statistical representation is considered and widely used as the generic case for thermal equilibrium of many uncorrelated or weakly-correlated systems.  However,  when effects such as ion-ion correlations, ion polarizability, finite size of ions, electrostriction and dielectric saturation of the solvent cannot be ignored, the GCS model becomes less reliable in describing the features of the EDL\;\cite{doi:10.1002/9780470142806.ch2, outhwaite1974higher,  moreira2000strong}. Furthermore, if the electrode surface is fractal and/or consisting of small confined geometries, the stochastic nature of state variables associated with the mobile ions for instance is also expected to deviate from Boltzmann's distribution profile\;\cite{garcia2011superstatistics}. 

The purpose of this study is to incorporate into 
the mean-field GCS model the Tsallis $q$-deformed exponential distribution \cite{tsallis2019beyond, GARCIAMORALES2004482, tsallis1988possible, beck2003superstatistics, beck2004superstatistics, abe2007superstatistics, garcia2011superstatistics} for the ion concentrations in the interfacial liquid (Sect.\;2). 
The same analysis can be applied in principle  to other approaches dealing with the EDL structure.  
The $q$-exponential function we are considering allows  the presence of random  fluctuations of ion concentrations around their mean values, which in turn provides a sort of a generalized Boltzmann factor. With this approach, one can still analyze a microscopic system in which correlations are expected as if they were absent\;\cite{garcia2011superstatistics}.  
 We provide analytical solutions for the charge density and differential capacitance from the boundary-value PB problem parameterized with the value of\;$q$ that can be attributed to the strength of elementary noise sources in the EDL structure. 
 A similar approach that  came to our attention during the writing of this paper was reported by Garcia-Morales et al.\;\cite{garcia2011superstatistics}, in which the focus was mainly on the $q$-dependence of the PB equation for describing counterion concentration around a charged surface.  
 However, this study is further concerned with  the overall   capacitive performance of the EDL structure by taking into account the series combination of the Helmholtz capacitance and the diffuse layer capacitance for a symmetric electrolyte,  which was not provided in ref.\;\cite{garcia2011superstatistics}.  We also evaluate the effect of the values of $q$ (taken between -1 to 1) on the EDL capacitance vs. voltage profiles which allowed to capture both the traditional inverse bell-shaped curves for aqueous solutions and bell-like curves measured for ionic liquids.

\section{The Gouy-Chapman-Stern Model}
\label{sec:1}
 The standard model to describe the equilibrium EDL structure at the junction metal/electrolyte (dilute solutions) is the GCS mean-field model which can quantitatively explain most experimental results. 
It consists of  subdividing the EDL into (i) a first sheet of uniform constant electronic charges on the metal surface, (ii) a charge-free inner compact layer or Helmholtz layer (also known as Stern layer) of a few angstrom in width and constant charge, and (iii) an outer, semi-infinite  diffuse layer (also known as Gouy-Chapman layer)  that extends into the  bulk electrolyte and contains anions and cations of a certain distribution\;\cite{OLDHAM2008131} (see Fig.\;\ref{fig1}).  Ions are assumed to be ideal point charges in local thermodynamic equilibrium within an isolated planar interface, and the  solvent is assumed to be a dielectric continuum  \cite{biesheuvel2010nonlinear}. 
There are many subsequent modified and refined versions of the GCS model that apply for instance corrections for ion crowding or dielectric saturation \cite{bazant2009towards, biesheuvel2007counterion, kornyshev2007double}, but here we use the simplest treatment for one electrode only that captures good enough the essentials.  Only physical adsorption in the diffuse  layer is considered to compensate the electronic charges, but no Faradaic charge-transfer reactions and/or ion adsorption are taken into account.

In the diffuse  layer part of the GCS's model,  the electrolyte is treated  using the mean-field   PB model. For dilute solutions with negligible correlations, the ions' concentrations  at a perpendicular distance  $ x \geqslant 0$ from a large lateral surface area metal electrode (one-dimensional problem in which the edge effects are ignored) are assumed to be related to the mean electrostatic potential $\psi_x$  by the  continuous Boltzmann distribution function:
 \begin{equation}
C_{x_i} = C_{x_i}^0 \exp \left( -\frac{z_iF\psi_x}{RT} \right)
\label{eq1}
\end{equation} 
where $C_{x_i}^0$ and  $z_i$  are the average ion concentration (for $\psi_x=0$) and  charge number of species $i$ (cation and anion), respectively. The term $RT/F=k_BT/e$ (with $R$  the ideal gas constant, $F$  the Faraday's constant, $k_B$  the Boltzmann's constant, $e$  the elemental electronic charge, and $T$  the absolute temperature) is the thermal voltage.
The molar electrical interaction energy in the Boltzmann factor takes into account the term $z_iF\psi_x$ only,  and other contributions are ignored in this study.  The approximation of $C_{x_i}$ by a  Boltzmann distribution is a standard tool used to evaluate many other thermodynamic properties of systems, but  is valid for some ideal conditions only, i.e. elastic collision between particles, particles are identical and independent of one another, etc. It assumes that the ratio of equilibrium ionic concentrations at two locations $x_i$ and $x_j$ in the solution, i.e. $C_{x_i}/C_{x_j}$, at which the mean electrostatic potentials are $\psi_{{x_i}}$ and $\psi_{{x_j}}$, respectively, is related to the work needed to carry an ion of charge $z e$ from $x_i$ to $x_j$, i.e, $-ze (\psi_{{x_i}}- \psi_{{x_j}})$, through the exponential Boltzmann factor given in Eq.\;\ref{eq1} \cite{boukamp1995linear}. 

Now  the  distribution of the mean electrostatic  potential $\psi_x$ in Eq.\;\ref{eq1} is governed by the Poisson equation  that connects the net volume charge density $\rho_x$ at the distance $x$ from the interface with the corresponding potential $\psi_x$ as:  
\begin{equation}
\frac{\mathrm{d}^2\psi_x}{\mathrm{d}x^2} = -\frac{\rho_x}{\epsilon}     
\end{equation} 
where $\epsilon$  is the uniform dielectric permittivity of the solvent in units of the vacuum dielectric constant $\epsilon_0$. 
The mean  charge  density $\rho_x$ can  be expressed as the sum  
$
\rho_x = \sum_i z_i F C_{x_i} 
$ over the ions species,  
which leads to the second-order nonlinear differential equation:
\begin{equation}
\frac{\mathrm{d}^2\psi_x}{\mathrm{d}x^2}  = -\frac{1}{\epsilon} \sum_i z_i F  C_{x_i}^0 \exp \left( -\frac{z_iF\psi_x}{RT} \right)
\label{eq3}
\end{equation} 
With the change of variables $p={\mathrm{d}\psi_x}/{\mathrm{d}x}$, such that ${\mathrm{d}^2\psi_x}/{\mathrm{d}x^2}=\mathrm{d}p/\mathrm{d}x=p\,\mathrm{d}p/\mathrm{d}\psi_x$, we rewrite Eq.\;\ref{eq3} as:
\begin{align}
\frac{1}{2} \left( \frac{\mathrm{d}\psi_x}{\mathrm{d}x} \right)^2 & = \frac{RT}{\epsilon} \sum_i  C_{x_i}^0\left[ \exp\left(\frac{-z_iF\psi_x}{RT} \right) \right] + B
\end{align}
The integration constant $B$ is determined using the boundary conditions $\psi_x\,(x\to\infty)=0$ (electroneutrality) and  $\mathrm{d}\psi_x/\mathrm{d}x=0$ (electrostatic field must vanish at the mid-plane) so that:
\begin{equation}
  \left( \frac{\mathrm{d}\psi_x}{\mathrm{d}x} \right)^2  = \frac{2RT}{\epsilon} \sum_i  C_{x_i}^0\left[ \exp\left(\frac{-z_iF\psi_x}{RT} \right) -1 \right]
  \label{eq5}
\end{equation}

\begin{figure}[h] \begin{center}
\includegraphics{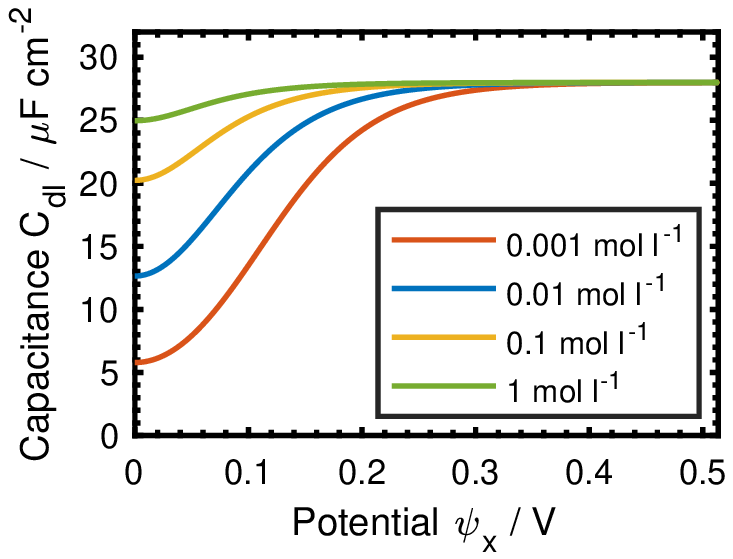}
\caption{Plots of double-layer capacitance (Eq.\;\ref{eq10}) as a function of the potential across the double layer for different values of $C_x^0$  with $C_{\text{H}}=\text{28}\,\mu\text{F\,cm}^{-2}$, $\epsilon=\text{80}\,\epsilon_0$, $z=\text{1}$, $T=\text{298}$\,K}
\label{fig2}
\end{center}
\end{figure}

If we consider the special case of a symmetric electrolyte (i.e. two ionic species; one positively charged and one negatively charged with $|z_i|=z$ and $C_{x_i}^0 = C_x^0, \forall i$), 
 we obtain the analytical solution of GCS model for a positively charged electrode as: 
\begin{equation}
\frac{\mathrm{d}\psi_x}{\mathrm{d}x} = -\sqrt{\frac{8RTC_x^0}{\epsilon}} \sinh \left ( \frac{zF\psi_x}{2RT} \right) < 0
\label{eq:6}
\end{equation} 
 in which physical arguments impose that the appropriate sign should be as shown. 
Note that Eq.\;\ref{eq:6} can be written in the dimensionless form 
$  {\mathrm{d} (\tilde{\psi}_x/2)}/{\mathrm{d} \tilde{x}} = - \sinh( \tilde{\psi}_x/2 )$, where $\tilde{\psi}_x = \psi / (RT/zF)$ and $\tilde{x} = x/\lambda_D$ with $\lambda_D = \sqrt{\epsilon RT/2 z^2 F^2 C_x^0}$ being the Debye screening length for the symmetric electrolyte. It can also be linearized in the limit of low potentials   leading to the Debye-H\"{u}ckel model.   
  With Eq.\;\ref{eq:6}  one can obtain the total net charge per unit surface $Q_x$ within the diffuse layer  as:
\begin{equation}
 \left( \frac{\mathrm{d}\psi_x}{\mathrm{d}x } \right)_{x=d} =-\frac{1}{\epsilon} \int\limits_{x=d}^{\infty} \rho_x \mathrm{d} x = -\frac{Q_x}{\epsilon}
\label{eq:7}
\end{equation}
 Combining Eq.\;\ref{eq:6} and Eq.\;\ref{eq:7} gives the total net charge within the diffuse layer as:
\begin{equation}
Q_x=  \sqrt{ {8 \epsilon RTC_x^0 } } \sinh \left( \frac{zF\psi_d}{2RT} \right)
\label{eq:8}
\end{equation}
To determine  the specific capacitance of the diffuse layer, the net charge (Eq.\;\ref{eq:8}) is  differentiated with respect to the potential, which gives the well-known nonlinear expression:
\begin{equation}
C_{\text{diff}}  =  \sqrt{\frac{2\epsilon z^2F^2 C_x^0}{RT}} \cosh \left( \frac{z F \psi_d}{2RT}  \right) 
\label{eq16}
\end{equation}
that depends on both  the potential $\psi_d$ and the properties of the electrolyte $C_x^0$ and $\epsilon$.  The expression for 
$C_{\text{diff}}$ can also be expressed in a dimensionless form as $\tilde{C}_{\text{diff}} = \lambda_D C_{\text{diff}} / \epsilon = \cosh( \tilde{\psi}_x/2 )$ \cite{bazant2004diffuse}.  
 The overall EDL capacitance of the electrode is then computed from the combination in series of the two capacitances arising from the Helmholtz inner layer (of constant capacitance $C_{\text{H}}=\epsilon/4\pi d$ with  $d$ being its thickness) and the diffuse layer, such that:
\begin{equation}
C_{\text{dl}}^{-1} = C_{\text{H}}^{-1} + C_{\text{diff}}^{-1}
\label{eq10}
\end{equation}
A plot of $C_{\text{dl}}$ vs. potential ($\psi_d>0$) for different values of the bulk concentrations $C_x^0$ is shown in Fig.\;\ref{fig2}. 
We took the values of 
$C_{\text{H}}=\text{28}\,\mu\text{F\,cm}^{-2}$, $\epsilon=\text{80}\,\epsilon_0$, $z=\text{1}$, and $T=\text{298}$\,K.   
At large voltages and for concentrated electrolytes, the overall capacitance is determined by  that of the Helmholtz layer, independently of the ion concentrations, whereas as the voltage is decreased and for very dilute electrolytes, the contribution of the diffuse layer becomes  more important. For  $\psi_d=0$, we have the reciprocal of the overall capacitance $C_{\text{dl}}^{-1} = C_{\text{H}}^{-1} + \left({{2\epsilon z^2F^2 C_x^0}/{RT}} \right)^{-0.5} $.

\section{The Generalized Boltzmann Factor}
\label{sec:3}

The Boltzmann  distribution function, as mentioned above, provides a valid approximation only under certain assumptions for  equilibrium thermodynamic \cite{hasegawa2007non, tsallis2004should, gray2018nonlinear}. While  it allows a relatively accurate  description of macroscopic systems in which a very large number of stochastic events take place, when one goes to smaller scales for instance, such a description breaks down \cite{garcia2010fluctuation}. 
  It is expected that at  small dimensions,   fluctuations of some intensive quantity such as temperature or pressure or chemical potential  for example  with respect to that of the surrounding reservoir, and contributions from the surface energy would play crucial roles in determining the state of the system under consideration \cite{bonnet2012first}. The system  in this case is in stationary quasi-thermodynamic equilibrium with the reservoir. This situation is closely related to the case of the EDL structure  within small sizes in the nanopores of porous electrodes. The effects of fluctuations and long-range interactions are actually more prominent  at this   level where the number of particles can be much smaller than Avogadro's number.  
  
  Moy et al. \cite{moy2000tests} have confirmed that some physical quantities,  such as the force on a test ion,  computed using the PB approximation deviates from the predictions with those of Brownian dynamics simulations (using the Langevin equation) in which  individual ions are treated explicitly. The difference widens as the distance of the ion from the channel is lower than the Debye's length.  The same discrepancies on the conductance and concentration profiles in cylindrical channels and a potassium channel were reported between the Poisson–Nernst–Planck theory (commonly used for non-equilibrium ion transport problems)  vs. the Brownian dynamics \cite{corry2000tests}. 
 In another study by H{\"a}rtel et al. \cite{hartel2015fundamental},  DFT and MD simulations were   applied to the EDL structure and  showed very good agreement for the ions concentration profiles, but clear deviations  from those computed with the PB theory are observed. 

Motivated by these observations,  the purpose of this study is to evaluate the behavior of the EDL structure using a more general Boltzmann factor as proposed by Tsallis \cite{tsallis1988possible}, and later extended by Beck and Cohen \cite{beck2003superstatistics, beck2004superstatistics, abe2007superstatistics} and others (see ref.\;\cite{tsallis2019beyond}   and references within). 
These  statistical mechanics  approaches have been introduced as a way to systematically describe complex statistical systems that behave like the superposition of many Boltzmann distributions, which makes the single Boltzmann distribution to be nothing but a special limiting case. Such systems usually involve long-range interactions, non-Markovian memory effects and anomalous diffusion; examples and applications can be found in different disciplines of science and engineering  \cite{gell2004nonextensive, abe2001nonextensive}. 
 At the microscopic level,    systems relax toward thermodynamic equilibrium (or more appropriately  stationary nonequilibrium \cite{hanel2011generalized}) following a single exponential decay, type Boltzmann, i.e. the probability of finding the system at some  specific state of energy $\epsilon$ is proportional to $\exp(-\beta \epsilon)$ where $\beta$ is a local inverse temperature. At a larger scale, the local variable $\beta$ (other variables can be considered) is not well-defined and   experiences fluctuations  following a certain probability distribution function $f(\beta)$, which depends on the underlying dynamics of the system and is a priori unknown \cite{hanel2011generalized}. 
 From Beck and Cohen \cite{beck2003superstatistics},  the generalized Boltzmann factor can be written as the integral over all possible fluctuating inverse temperatures $\beta$ as:  
\begin{equation}
B(E) = \int\limits_0^{\infty} f(\beta) \exp({-\beta E})  \mathrm{d}\beta 
\label{eq:BE}
\end{equation}
For most physical systems a distribution where the random variable $\beta$ is nonnegative is needed.  In particular, if   
 $f(\beta)$ is assumed to follow a gamma probability distribution function which arises naturally for a fluctuating environment with a finite number of degrees of freedom, such that:
\begin{equation}
f(\beta) = \frac{1}{b \Gamma(c)} \left({\beta}/{b}\right)^{c-1} \exp \left(-{\beta}/{b} \right)
\end{equation}
 where $c$ and $b$ are positive parameters, 
     Eq.\;\ref{eq:BE} leads to the generalized Boltzmann factor in Tsallis statistics  \cite{beck2003superstatistics, tsallis2019beyond, tsallis1988possible, tsallis1999nonextensive,  tsallis2004should, abe2001nonextensive}:
\begin{equation}
B(E) = (1+bE)^{-c} = \left[ 1- (1-q)\beta_0 E \right]^{{1}/{(1-q)}} = \exp_q({-\beta_0 E})
\label{eq:Tsallis}
\end{equation}
with $1/(q-1)=c$, 
$\beta_0= \int_0^{\infty} \beta f(\beta) \mathrm{d}\beta 
= bc$ is the average of the fluctuating $\beta$, and  
  $\exp_q(y)$ denotes the $q$-exponential function. The parameter $q$ is a real number that characterizes the system's statistics, and is defined by the ratio of standard variation and mean of the distribution $f(\beta)$  ($q=1$ if there are no fluctuations) \cite{beck2003superstatistics}. Note that 
  (i) for  $q<1$, $\exp_q(y)=0$ for $y<-1/(1-q)$ and $\exp_q(y)=\left[1+(1-q)y \right]^{1/(1-q)}$ for $y\geqslant -1/(1-q)$, 
  (ii) for $q=1$, $\exp_q(y)=\exp(y)$ for $\forall y$, and 
  (iii) for $q>1$, $\exp_q(y)= \left[1+(1-q)y \right]^{1/(1-q)}$ for $y < 1/(q-1)$ \cite{abe2001nonextensive}.

\section{The Extended Gouy-Chapman-Stern Model}
\label{sec:2}

\begin{figure}[h] \begin{center}
\includegraphics{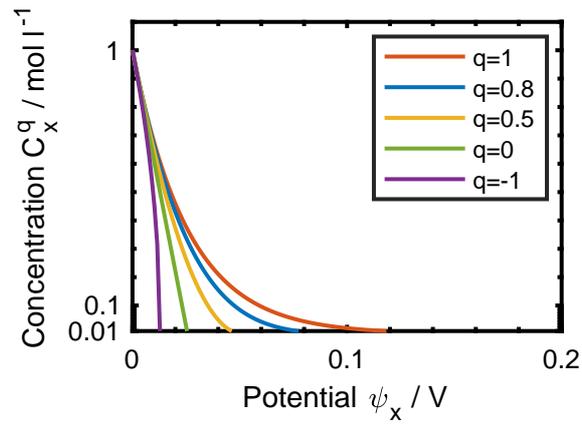} 
\caption{Plots of 
 ion concentration (Eq.\;\ref{eqCxq}) 
as a function of the potential across the double layer  for different values of $q$ and with $C_{x}^0=1\,\text{mol\,l}^{-1}$, $z=\text{1}$, $T=\text{298}$\,K. The straight lines 
  with the slope of ${-z F}/{R T \ln(10)}$ are for $q=1.0$ only, otherwise the decay-plots follow power-law  profiles.}
\label{fig3}
\end{center}
\end{figure}

By applying generically the generalized Boltzmann factor given by Eq.\;\ref{eq:Tsallis} to the concentrations of charged ions in the EDL structure (instead of Eq.\;\ref{eq1}), we write  the $q$-exponential relation \cite{GARCIAMORALES2004482}:  
\begin{equation}
C_{x_i}^q = C_{x_i}^0 \exp_q \left(-\frac{z_i F \psi_x}{R T} \right)
\label{eqCxq}
\end{equation}
Again, this means  that the fluctuations of the mean-field value of the ion concentrations are assumed to follow a gamma distribution with the parameter $q=1+1/c$ representing  the extent of these fluctuations. The adequate determination of the statistical distribution of such ions concentration is very difficult to obtain. If $q=1$, the traditional Boltzmann factor given in Eq.\;\ref{eq1} is immediately recovered from Eq.\;\ref{eqCxq}.  
Plots of $C_{x_i}^q$ 
 as a function of potential $\psi_x$ for different values of $q \leqslant 1$ are shown in Fig.\;\ref{fig3}. In a semi-logarithmic scale, it is only the  Boltzmann distribution that would result in a straight line with slope $-zF/RT \ln(10)$. Otherwise,  Decreasing the value of $q$ from unity implies steeper decrease of the ionic concentration following a power-law profile as shown in the figure.  
This means that contrary to the exponential decay in which subsystems are uncorrelated, values of $q$ different from one implies some form of internal correlations between subsystems.

The $q$-modified PB model becomes then:
\begin{equation}
\mathrm{d} \left(\frac{ \mathrm{d} \psi_x}{\mathrm{d} x} \right)^2  =  - \frac{2 F}{\epsilon} \sum_i C_{x_i}^0 z_i \exp_q \left(- \frac{z_i F \psi_x}{R T} \right) \mathrm{d} \psi_x
\end{equation}
which leads after integration to: 
\begin{align}
\left(\frac{\mathrm{d} \psi_x}{\mathrm{d} x} \right)^2 &=  \frac{2 R T}{\epsilon (2-q)} \times \nonumber \\ 
&  \sum_i C_{x_i}^0 \left[ \left(1 - (1-q) \frac{z_i F \psi_x}{R T} \right) 
\exp_q \left(- \frac{z_i F \psi_x}{R T} \right)  -1 \right]
\label{eq17}
\end{align}
We verify that for $q=1$, Eq.\;\ref{eq17}  readily simplifies to the expression given in Eq.\;\ref{eq5}. 
 For the case of a symmetric 1:1 electrolyte, we obtain: 
\begin{align}
\left(\frac{\mathrm{d} \psi_x}{\mathrm{d} x} \right)^2 &  =  \frac{4 R T C_x^0}{\epsilon (2-q)} \times \nonumber \\
 & \left[ \cosh_q \left(\frac{z F \psi_x}{R T} \right) + 
(1-q) \left(\frac{z F \psi_x}{R T} \right) \sinh_q \left(\frac{z F \psi_x}{R T} \right) -1 \right]
\end{align}
where the $q$-hyperbolic functions $\cosh_q$ and $\sinh_q$ are defined respectively as \cite{borges1998q}:
\begin{subequations}
\begin{eqnarray}
2 \cosh_q (y) &= & {\exp_q (y) + \exp_q (-y)}  \\
2 \sinh_q (y) & = & { \exp_q (y) - \exp_q (-y)}
\end{eqnarray}
\end{subequations}
Now using the $q$-identities: 
\begin{subequations}
\begin{eqnarray}
\cosh_q \left(y \right) & = & 2 \sinh_{2q-1}^2 \left({y}/{2} \right) + \exp_{2q-1} \left[-(1-q)  {y^2}/{2} \right]   \\
\sinh_q \left(y \right) & = & 2 \sinh_{2q-1} \left( {y}/{2} \right) \cosh_{2q-1} \left( {y}/{2} \right) 
\end{eqnarray}
\end{subequations}
we obtain the following expression for the electric field: 
\begin{eqnarray}
\frac{\mathrm{d} \psi_x}{\mathrm{d} x} & = & - \sqrt{\frac{8 R T C_x^0}{\epsilon (2-q)}}  \sinh_{2q-1} \left(\frac{z F \psi_x}{2 R T} \right) h_q \left(\frac{z F \psi_x}{R T} \right)
\label{eq20}
\end{eqnarray}
where the function $h_q(y)$ is given by:
\begin{equation}
h_q (y) =  \left[1 + (1-q) y \coth_{2q-1} \left(\frac{y}{2} \right) - 
\frac{1- \exp_{2q-1} \left(-(1-q) \frac{y^2}{2} \right)}{2 \sinh_{2q-1}^2 \left(\frac{y}{2} \right)}\right]^{0.5}
\end{equation}
 It is clear that when $q\to 1$, the classical relation given by Eq.\;\ref{eq:6} for the electric field is immediately recovered. 
 We can write Eq.\;\ref{eq20} in a dimensionless form as $ {\mathrm{d} (\tilde{\psi}_x/2)}/{\mathrm{d} \tilde{x}_q}   =  -  \sinh_{2q-1}  ( \tilde{\psi}_x/2  ) h  ( \tilde{\psi}_x  )$ where $\tilde{x}_q$ is the normalized dimension $x$ with respect to a modified Debye  length, $\lambda_D^q =\lambda_D \sqrt{2-q}$, which depends on the properties of the liquid as well as the extent of the fluctuations through the parameter $q$.  $\lambda_D^q $ is larger than $\lambda_D$ for $q<1$ which implies that the scale over which the electrolyte screens the surface charge is larger with the presence of fluctuations.  
 Finally, the $q$-parameterized net charge $Q_{x}^{q}$ in the diffuse layer using the generalized Boltzmann factor is found as: 
\begin{equation} 
Q_{x}^{q}  =  \sqrt{\frac{8 \epsilon R T C_x^0  }{2-q}} 
\sinh_{2q-1} \left(\frac{z F \psi_d}{2 R T} \right) h_q \left(\frac{z F \psi_d}{R T} \right)
\end{equation}
and its associated diffuse capacitance is found numerically from $C_{\text{diff}}^{q} = \partial Q_{x}^{q} / \partial \psi_x$. 
 The overall EDL capacitance $C_{\text{dl}}^{q}$ is then computed as per Eq.\;\ref{eq10}, i.e.:
 \begin{equation}
(C_{\text{dl}}^q)^{-1} = C_{\text{H}}^{-1} + (C_{\text{diff}}^{q})^{-1}
\label{eq23}
\end{equation}
 The total charge $Q_{\text{dl}}^{q}$ is obtained from the voltage integral of $C_{\text{dl}}^{q}$. 
 Plots of  $C_{\text{dl}}^{q}(\psi_x)$ for 
 different values of $q$ for a constant bulk concentration of $C_{x}^0=10^{-3}\,\text{mol\,L}^{-1}$,  
  are shown in Fig.\;\ref{fig4}.   The capacitance at negative voltages exactly duplicate the capacitance at positive voltages shown in the figure.

\section{Discussion}

\begin{figure}[h] \begin{center}
\includegraphics{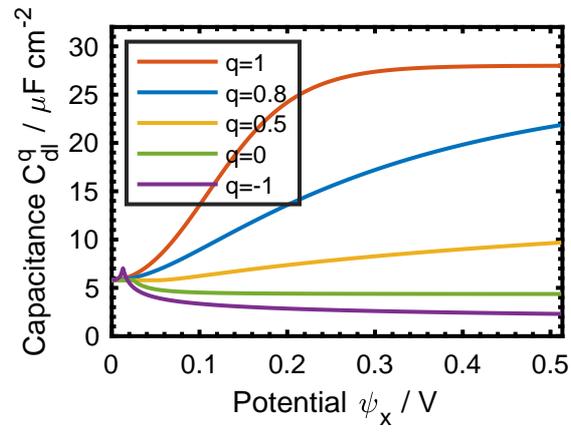}
\caption{Plots of double-layer capacitance (Eq.\;\ref{eq23}) as a function of the potential across the double layer for  
  different values of $q$ (from -1 to 1) with $C_{x}^0=10^{-3}\,\text{mol\,l}^{-1}$,  $C_{\text{H}}=\text{28}\,\mu\text{F\,cm}^{-2}$, $\epsilon=\text{80}\,\epsilon_0$, $z=\text{1}$ and $T=\text{298}$\,K}
\label{fig4}
\end{center}
\end{figure}

From Fig.\;\ref{fig4}, it is evident that incorporating the generalized Boltzmann factor 
 for the ions concentrations into the mean-field GCS model affects greatly the capacitive performance of the EDL structure. It is understood that other electrical characteristics are also impacted, but the focus of the discussion here is on the differential EDL  capacitance, and thus energy storage applications. 
The strength of the fluctuations increases as the value of $q$ deviates further from unity at which the traditional Boltzmann factor is recovered. 
In other words, when the parameter $q$ is less than one, it implies that there are some forms of hidden correlations within the stationary system in quasi-thermodynamic equilibrium, which means that its constituting subsystems are  not statistically independent anymore as it is the case in Boltzmann theory \cite{hanel2011generalized}.  In practice, however, the determination of an experimental value for $q$ and the particular form of correlations between the dynamic subsystems is not trivial as mentioned above. 

Fig.\;\ref{fig4} shows also that the overall capacitance exhibits a local minimum at the origin for different values of $q>0$, and an asymptotic increase towards the value of  $C_{\text{H}}=\text{28}\,\mu\text{F\,cm}^{-2}$ at higher  polarization.  Otherwise, between these two limiting cases, the shapes of the profiles for $0<q<1$ are in concordance with the shapes obtained with the traditional GCS model for which $q=1$ (compare   with Fig.\;\ref{fig2}). 
Now because of the interdependence of internal subsystems for $q<1$, this results is an extra contribution to the amount of entropy of the system, i.e. the entropy $S_{q}(A+B) \geqslant S_q(A) + S_q(B) $ of two subsystems $A$ and $B$ \cite{tsallis2004should}. This non-additivity of entropy is the basis of a non-extensive Tsallis statistics. 
   Tsallis proposed the  generalized  entropy  
$S_q 
= -k \sum_i p_i^q \ln_q(p_i)
$ \cite{tsallis1988possible, tsallis1999nonextensive,  tsallis2004should, abe2001nonextensive}, where $k$ is a positive constant, $q\neq 1$, 
 and the quantities $p_i=p(E_i)$ represent the probabilities for the occurrence of the $i^{\text{th}}$ microstate and satisfy $\sum_{i} p_i=1$. The function $\ln_q(y)=(y^{1-q}-1)/(1-q)$ denotes the $q$-logarithm, inverse of the $q$-exponential (i.e. $\ln_q[\exp_q(y)]=\exp_q[\ln_q(y)]=y$) \cite{tsallis2004should}. In the limit of $q\to 1$, one recovers the classical form of Boltzmann-Gibbs entropy 
$S_1 = -k_B \sum_{i} p_i \ln (p_i)$ 
 where $k=k_B$ is the Boltzmann constant \cite{tsallis1988possible}. 
Such additional amount of entropy, which is related to the value of $q$, affects negatively the net density of charge in the diffuse layer of the EDL structure compared to the traditional case when $q=1$. This in turn lowers the overall EDL capacitance as depicted in Fig.\;\ref{fig4}. 
 
 What is also remarkable from these results is the bell-shaped-like trend of the total capacitance when $q$ takes values less than zero. A similar  profile is observed for metal/ionic liquid double layer capacitance which exhibit a local maximum at the point of zero charge \cite{kornyshev2007double}. In Fig.\;\ref{fig4}, we show the example for $q=-1$ which is typical for all negative values of $q$.  The capacitance shows a peaking value and reverse of curvature close to the origin at $\psi_x=0$, and the same is observed when the capacitance is plotted vs. charge. It is well known  that the non-linear PB theory of points ions is unable to display a change of curvature observed experimentally in ionic liquids or highly concentrated electrolytes. 
 This change of curvature can be, nonetheless, predicted in the so-called primitive model (in which solvent effects are considered implicitly) by including ion correlations and ionic excluded volume effects. This has been shown by numerous DFT and  integral equations calculations, as well as Monte Carlo, and Brownian dynamics and molecular dynamics simulations. The change of curvature of the differential capacity can be also observed in more detailed atomistic models in which solvent particles are taken into account explicitly.

Finally, the authors recognize that the main limitation of this theoretical description is the lack of a proper physical meaning for the parameter $q$ in terms of a microscopic model. 
The analysis presented here is unable to provide an effective description of ion correlations (i.e., the Coulombic attraction of charged particles of opposite valence and the Coulombic repulsion of charged particles with the same valence) and ionic excluded volume effects (i.e., the fact that two ions cannot overlap due to its finite size), but this is the subject of ongoing investigations. 
 A measurable atomistic model supporting the physical reality of the parameter $q$ will follow to validate   the use of non-extensive Tsallis statistics in EDL systems.
 In this sense, the parameter $q$ can be viewed as mathematically well-fit to capture the change of curvature of the differential capacity, as it is the case for instance of fractional integro-differential equations for capturing anomalous time-domain and frequency-domain features in several physicochemical systems \cite{fracorderreview,memoryAPL,energy2015,orgElectronics}.

\section{Conclusion}

In this paper, we analyzed the effect of  superposing fluctuations onto the mean-value Boltzmann distribution of ion concentrations in the GCS model on the capacitive properties of the EDL structure. This was done  via the $q$-exponential function  which allows to embrace a spectrum of empirical processes connected to the degree of fluctuations in the intensive inverse temperature parameter. This compact and efficient approach provides an extended version to the classical GCS model specifically for the description of the EDL at scales in which correlation between subsystem are possible.  The analytical expression for the EDL capacitance as a function of voltage and  ion concentration, and parameterized with the index $q<1$, predicts lower values than its GCS counterpart (i.e. for $q=1$) which has been explained by super-additive entropy feature in Tsallis statics.  Additionally, for values of $q$ less than zero, the model reverses the general trend of capacitance at a metal/electrolyte junction towards that of a bell-shaped-like profiles which are observed in ionic liquid EDL structure. Further investigations in this direction with comparison with atomistic models are being developed for future contributions.

\section*{Acknowledgement}
   
We thank Mathijs Janssen for useful discussions and comments on an earlier version of the manuscript.

\section*{References}

%

\bigskip

\biboptions{numbers,sort&compress}

\newpage \clearpage
 
\newpage

\newpage

\newpage

 \end{document}